\RequirePackage{lineno} 
\documentclass{nature}
\usepackage{xcolor}

\usepackage{caption}
\usepackage{multirow}
\usepackage{amsmath}
\usepackage{CJK}
\usepackage{bm}
\usepackage{braket}
\usepackage{amsmath}

\usepackage{graphicx}
\usepackage{dcolumn}
\usepackage{stmaryrd}
\usepackage{latexsym}
\usepackage{amssymb}
\usepackage{amsfonts}
\usepackage{fancybox}
\usepackage{color}
\usepackage{verbatim}
\usepackage{physics}
\usepackage{bbold}

\usepackage{amsmath}
\usepackage{bm}
\usepackage{CJK}
\usepackage{braket}
\usepackage{float}
\usepackage{amssymb}
\usepackage{siunitx}

\title{Correlated topological flat bands in rhombohedral graphite}

\author{Hongyun Zhang$^{1,\dagger}$, Qian Li$^{1,\dagger}$, Michael G. Scheer$^{2}$, Renqi Wang$^{1}$, Chuyi Tuo$^3$, Nianlong Zou$^{1}$, Wanying Chen$^1$, Jiaheng Li$^1$, Xuanxi Cai$^{1}$, Changhua Bao$^1$, Ming-Rui Li$^3$, Ke Deng$^1$, Kenji Watanabe$^4$, Takashi Taniguchi$^5$, Mao Ye$^6$, Peizhe Tang$^{7,8}$, Yong Xu$^{1,9}$, Pu Yu$^{1,9}$, Jose Avila$^{10}$, Pavel Dudin$^{10}$, Jonathan D. Denlinger$^{11}$, Hong Yao$^{3}$, Biao Lian$^{2}$, Wenhui Duan$^{1,3,9}$ \& Shuyun Zhou$^{1,9,*}$}

\usepackage{graphicx}
\makeatletter
\let\saved@includegraphics\includegraphics
\AtBeginDocument{\let\includegraphics\saved@includegraphics}

\makeatother

\begin{document}

	\maketitle
	
	\begin{affiliations}
		
		\item State Key Laboratory of Low-Dimensional Quantum Physics and Department of Physics, Tsinghua University, Beijing 100084, P. R. China
		\item Department of Physics, Princeton University, Princeton, New Jersey 08544, USA
		\item Institute for Advanced Study, Tsinghua University, Beijing 100084, P. R. China
		\item Research Center for Functional Materials, National Institute for Materials Science, 1-1 Namiki, Tsukuba 305-0044, Japan
		\item International Center for Materials Nanoarchitectonics, National Institute for Materials Science, 1-1 Namiki, Tsukuba 305-0044, Japan
		\item Shanghai Synchrotron Radiation Facility, Shanghai Advanced Research Institute, Chinese Academy of Sciences, Shanghai 201210, P. R. China
		\item School of Materials Science and Engineering, Beihang University, Beijing 100191, P. R. China
		\item Max Planck Institute for the Structure and Dynamics of Matter, Center for Free Electron Laser Science, Hamburg 22761, Germany
		\item Frontier Science Center for Quantum Information, Beijing 100084, P. R. China
		\item Synchrotron SOLEIL, L’Orme des Merisiers, Saint Aubin-BP 48, 91192 Gif sur Yvette Cedex, France
		\item Advanced Light Source, Lawrence Berkeley National Laboratory, Berkeley, California 94720, USA\\
		$\dagger$ These authors contributed equally to this work.\\
		*Correspondence should be sent to syzhou@mail.tsinghua.edu.cn.
	\end{affiliations}

	\begin{abstract}
		
Flat bands and nontrivial topological physics are two important topics of condensed matter physics. With a unique stacking configuration analogous to the Su-Schrieffer-Heeger (SSH) model, rhombohedral graphite (RG) is a potential candidate for realizing both flat bands and nontrivial topological physics. 
Here we report experimental evidence of topological flat bands (TFBs) on the surface of bulk RG, which are topologically protected by bulk helical Dirac nodal lines via the bulk-boundary correspondence. Moreover, upon {\it in situ} electron doping, the surface TFBs show a splitting with exotic doping evolution, with an order-of-magnitude increase in the bandwidth of the lower split band, and pinning of the upper band near the Fermi level. These experimental observations together with Hartree-Fock calculations suggest that correlation effects are important in this system. Our results demonstrate RG as a new platform for investigating the rich interplay between nontrivial band topology, correlation effects, and interaction-driven symmetry-broken states.
	\end{abstract}

	\newpage
	
	\renewcommand{\thefigure}{\textbf{Fig. \arabic{figure} $\bm{|}$}}
	\setcounter{figure}{0}
	
The stacking order as well as the twisting angle of graphene layers provides a control knob for inducing flat band and emergent phenomena, such as superconductivity \cite{PabloNat2018}, correlated insulation \cite{PabloMott2018}, quantum anomalous Hall effect \cite{YoungQAHSci2020,JuLQAHE}, and fractional quantum anomalous Hall effect \cite{JuL2023FQAHE}.  In few-layer rhombohedral graphene with ABC stacking order, a wide range of intriguing properties have been reported in few-layer flakes with varying thickness as well as in moir\'e superstructures. For example, the bandgap and transport properties are highly tunable by an electric field \cite{TaruchaNatNano2009,HeinzTriGNP2011,LauNatPhys2011}. Interaction-driven correlated phenomena and competing symmetry-broken states, such as Mott insulation, superconductivity, ferromagnetism, and ferroelectricity, have been reported in trilayer flakes and moir\'e superlattices \cite{WangFTrilayerNP19,WangFTrilayerNat19,WangFTrilayerNat19FM,YoungSC2021,YoungHalfMetalNat2021,JuLScience2022}, tetralayer and heterostructures \cite{PasupathyPNAS2021,ChenGRNatNano2023,ChenGRSOC2023}, and even thicker  rhombohedral graphene systems \cite{MishchenkoNat2020,JuLNatNano2023,JuLMultiferroicNat2023}. More interestingly, the quantum anomalous Hall effect \cite{JuLQAHE} and the fractional quantum anomalous Hall effect \cite{JuL2023FQAHE} have been recently discovered in pentalayer rhombohedral graphene moir\'e heterostructures.

\begin{figure*}
	\centering
	\includegraphics[width=17cm]{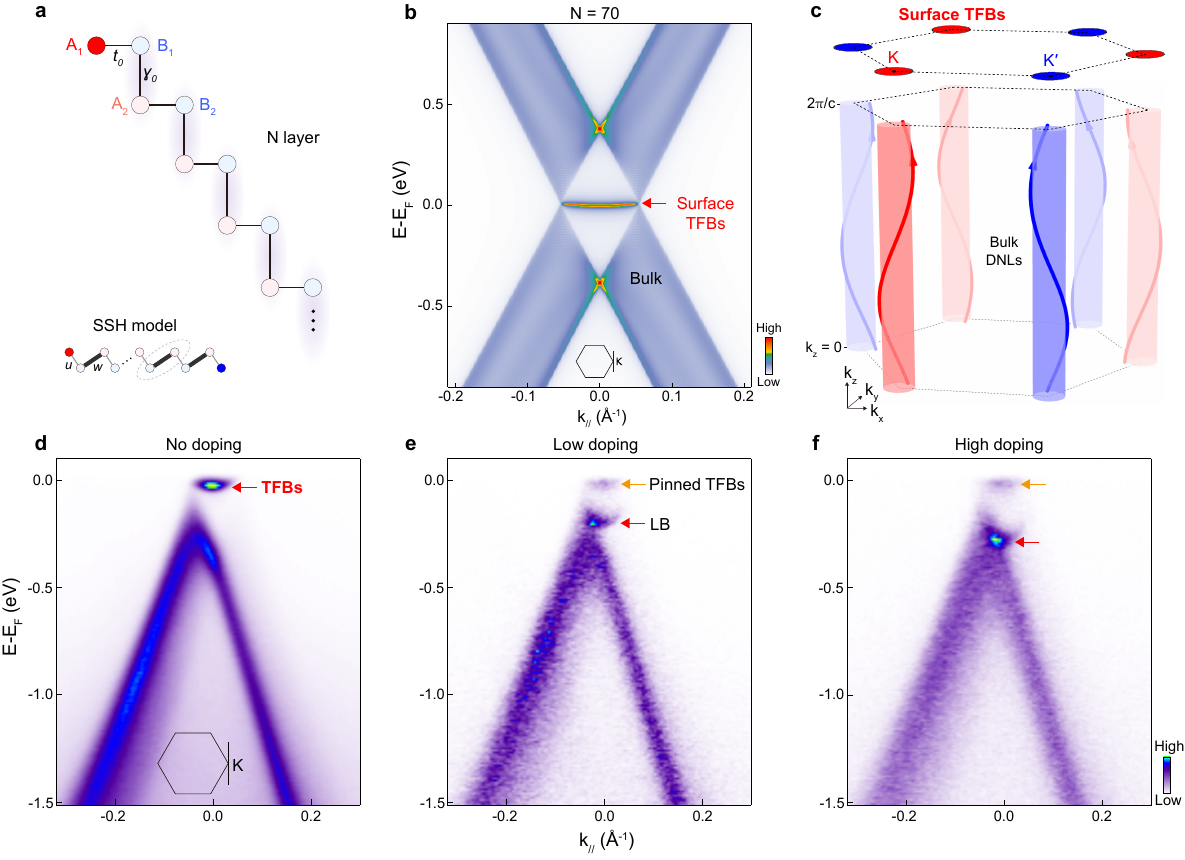}
	\caption{\textbf{Schematic illustration of RG, surface topological flat bands (surface TFBs) and bulk helical Dirac nodal lines (bulk helical DNLs), and experimental observations.} (\textbf{a}), Schematic illustration of the atomic structure of bulk RG (side view), which is analogous to the SSH model (inset in the lower left corner). (\textbf{b}), Calculated electronic spectral weight for RG with N = 70 layers. The surface TFBs are indicated by the red arrow, while the light-blue colors arise from the bulk conical bands. (\textbf{c}), Schematic illustration of the surface TFBs (red and blue ovals) and bulk helical DNLs (red and blue helixes). (\textbf{d}), ARPES dispersion image measured by cutting through the K point as indicated by the inset. (\textbf{e, f}), Dispersion images measured upon low and high electron doping, respectively. The splitting of the TFBs is indicated by red and orange arrows, and the lower split band is denoted by LB.}
	\label{fig1}
\end{figure*}

The origin of these interesting correlated phenomena is a band flattening effect that occurs near the K and K$^\prime$ points. The bands nearest the Fermi level $E_F$ exhibit a power law behavior, $E \approx \pm \lvert k\rvert^N$, where $E$ is the electron energy, $k$ is the two-dimensional momentum relative to the K or K$^\prime$ point, and $N$ is the number of graphene layers \cite{Koshino2009trigonal,MacDonaldPRB2010,StarkePRB2013,Ouerghi2015,BaoNL2017,SunJWNL2018FB,OuerghiPRB2018}. As the  parent compound of few-layer rhombohedral graphene, bulk rhombohedral graphite (bulk RG) is expected to have an even flatter band, which could further enhance instabilities toward interaction-driven symmetry-broken states. Moreover, its unique stacking sequence makes it a potential three-dimensional topological semimetal with $\mathbb{Z}_2$ topological charge \cite{Volovik_JETP2011}, making it a potential candidate for hosting both bulk topological physics and flat band, which can go beyond the physics of twisted bilayer graphene and few-layer rhombohedral graphene. In addition, exotic properties such as topological, chiral, or high-temperature surface superconductivity \cite{VolovikSCPRB2011,SerbynPRB2023} have also been predicted. It is therefore important to investigate if bulk RG exhibits correlation effect and nontrivial topological properties.

In this work, we report direct experimental evidence of surface topological flat bands (surface TFBs) with correlation effects in RG using angle-resolved photoemission spectroscopy (ARPES) with micrometer or nanometer scale beam size (MicroARPES or NanoARPES). Weakly dispersing surface TFBs are clearly observed near the K and K$^\prime$ points, along with corresponding bulk helical Dirac nodal lines (DNLs). Upon {\it in situ} electron doping, the surface TFBs in the K valley split into two bands, the lower of which (labeled as LB) increases in the bandwidth by an order of magnitude, while the upper one remains pinned near $E_F$. Self-consistent Hartree-Fock calculations reveal spontaneous flavor symmetry breaking and doping evolution of the lower band similar to that observed experimentally. However, the Hartree-Fock results do not fully explain the pinning of the upper band near $E_F$, suggesting that electron-electron correlation effects beyond mean-field theory are important for a full explanation.

\section{Observation of bulk helical Dirac nodes and surface flat bands}

Figure~1a shows a schematic sideview of RG. The stacking order is  analogous to zigzag graphene ribbons \cite{Steven_Nature2006} or the one-dimensional (1D) Su-Schrieffer-Heeger (SSH) topological model \cite{SSHmodel_PRL1979}. Here the in-plane momentum ($k$) kinetic term $\hbar v_F\lvert k\rvert$ ($v_F$ is the Fermi velocity) and out-of-plane hopping $\gamma_0$ are mapped  (up to a phase) onto the staggered hopping parameters $u$ and $w$ in the SSH model \cite{TautPRB2011} (see inset in Fig.~1a). In few-layer RG, each carbon sublattice hybridizes with orbitals directly above or below in the neighboring layers, forming conical subbands, while the unpaired orbitals in the outermost layers contribute to weakly dispersing surface flat bands near $E_F$. When increasing the layer number $N$, the surface flat bands become flatter and a higher density of states is expected at $E_F$ (see SI Appendix Fig.~S1), while the bulk bands evolve into Dirac cones. Figure~1b shows the calculated electronic spectral weight for bulk RG with N = 70 layers (see Methods for calculation details). Here, the conical regions with light-blue colors correspond to bulk bands, whose Dirac nodes form helixes along the out-of-plane momentum ($k_z$) direction with opposite chiralities near the K and K$^\prime$ points \cite{mcclure1969electron,Volovik_JETP2011,LinMFPRB2016} (see schematic illustration in Fig.~1c). These bulk helical DNLs have a quantized Zak phase of $\pi$ ($-\pi$) near the K (K$^\prime$) point when $\hbar v_F\lvert k\rvert<\gamma_0$, which guarantees the existence of topologically protected ``drumhead'' surface states, hereafter denoted as surface topological flat bands (surface TFBs), due to the bulk-boundary correspondence \cite{BalentsPRB2011,FuLNLSPRB2015,RappePRL2015,KawazoeWengPRB2015,FangZNLSCPB2016}. Bulk RG is therefore a model system for realizing a helical Dirac nodal line semimetal \cite{Volovik_JETP2011}, where the surface TFBs (indicated by the red arrow in Fig.~1b) naturally arise as a result of the bulk band topology and are protected by the combined inversion- and time-reversal ($\mathcal{PT}$) symmetry.

\begin{figure*}
	\centering
	\includegraphics[width=17cm]{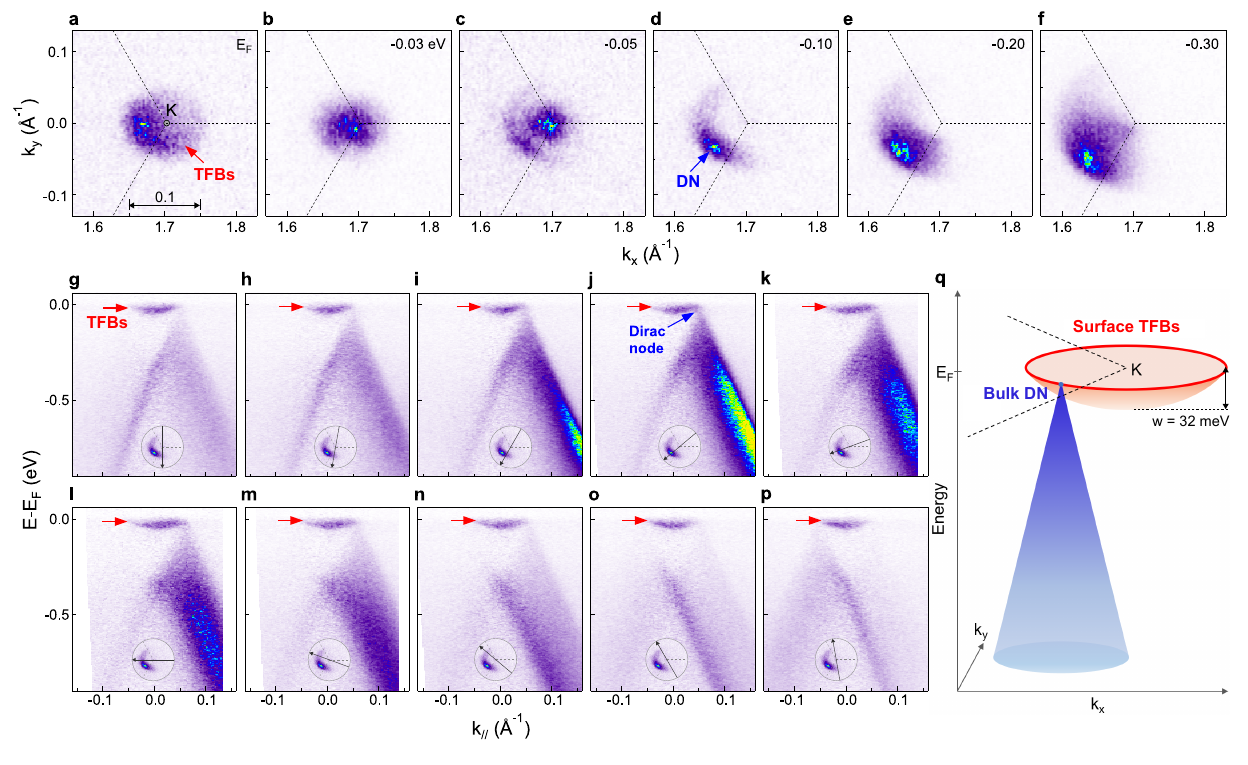}
	\label{Fig.2|}
	\caption{\textbf{Observation of surface TFBs and bulk Dirac node.} (\textbf{a-f}), ARPES-measured intensity maps at a few representative energies from $E_F$ to $\SI{-0.30}{\electronvolt}$ measured on sample S2. The photon energy used is $\SI{60}{\electronvolt}$, which corresponds to $k_z = 0.11c^*$ in the reduced BZ ($c^* = \frac{2\pi}{c}$, where $c=\SI{3.34}{\angstrom}$ is the out-of-plane lattice constant) for the K point. The red arrow indicates the TFBs and the blue arrow in (\textbf{d}) indicates the bulk Dirac node (bulk DN). (\textbf{g-p}), Dispersion images measured by cutting through the K point along different directions indicated by arrows in the insets. (\textbf{q}), Schematic summary of the experimental electronic structure, including the DN (blue color) and the TFBs near $E_F$ (red color).}
	\label{fig2}
\end{figure*}

The RG samples were prepared by mechanical exfoliation, and the ABC stacking was confirmed by spatially-resolved Raman spectroscopy and NanoARPES intensity maps (see SI Appendix Fig.~S2,3). Figure~1d shows an overview of the characteristic ARPES dispersion image measured through the K point and the doping evolution. The surface TFBs near $E_F$ (indicated by the red arrow) and the conical bulk states are clearly observed. The surface TFBs show exotic evolution upon electron doping, with a splitting of the surface TFBs indicated by orange and red arrows in Fig.~1e. An increase in the bandwidth is observed for the lower band at high electron doping, while the upper band is pinned near $E_F$ (Fig.~1f). The origin of the surface TFBs and their correlated doping evolution are the main focus of this work.

To reveal the details of the surface TFBs and bulk states, Figure~2 shows the experimental electronic structure of bulk RG measured at a photon energy of $\SI{60}{\electronvolt}$. The Fermi surface map in Fig.~2a shows a circular-shaped pocket centered at the K point, which decreases in size and shrinks into a dot when moving down to $\SI{-0.03}{\electronvolt}$ (Fig.~2b), indicating small bandwidth of TFBs. In addition, when moving down to $\SI{-0.10}{\electronvolt}$ (Fig.~2c,d), a bright spot is observed at the lower-left corner in Fig.~2d (indicated by the blue arrow), which gradually expands in size at lower energies, indicating a conical behavior (Fig.~2e,f). 

The Dirac cone and TFBs are more clearly resolved in dispersion images measured by cutting through the K point along various azimuthal angles (Fig.~2g-p). The Dirac cone is most clearly observed with the sharpest dispersion when cutting through the bright spot in Fig.~2j, suggesting that it likely originates from the bulk band at a specific $k_z$ value corresponding to the probe photon energy of $\SI{60}{\electronvolt}$, while the weaker and broadened Dirac cones originate from intensity tail at other $k_z$ values (see SI Appendix Fig.~S4). The Dirac cone is connected to the edges of the flat band near $E_F$, which is observed in all cuts through the circular pockets at $E_F$ (indicated by red arrows in Fig.~2g-p). A detailed analysis of the dispersion shows that it has a bandwidth of $32 \pm  \SI{7}{\milli\electronvolt}$ and a momentum range of $\Delta k = 0.10 \pm \SI{0.02}{\per\angstrom}$ around the K point (see SI Appendix Fig.~S5 for more details). According to the theoretical models \cite{TautPRB2011,HeikkilaJLT2018}, the momentum radius $p_0$ of the flat band in RG is defined by $p_0=\gamma_0 /\hbar v_F=2\gamma_0 /(\sqrt{3}at_0)$ where $t_0$ is the nearest-neighbor hopping and $a$ is the lattice constant. Using the values $t_0 = \SI{3.16}{\electronvolt}$, $\gamma_0 = \SI{0.39}{\electronvolt}$ (ref. \cite{Koshino2009trigonal}), and $a = \SI{2.46}{\angstrom}$ gives a momentum range of $\Delta k = 2p_0 = \SI{0.11}{\per\angstrom}$, in agreement with the experimental result. Therefore, the experimental electronic structure in Fig.~2 can be summarized by the TFBs near $E_F$ (red color) and a connecting Dirac cone (blue color), as schematically illustrated in Fig.~2q.

\begin{figure*}
	\centering
	\includegraphics[width=17cm]{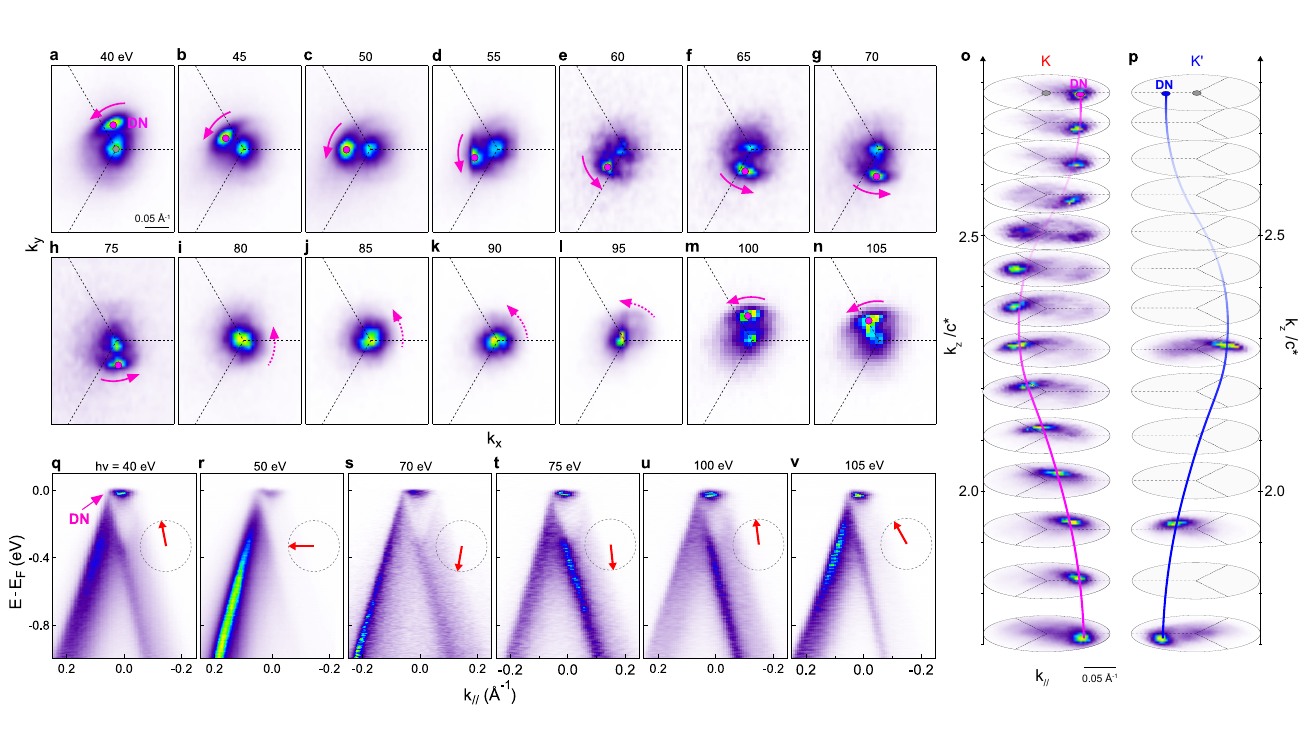}
	\caption{\textbf{Observation of bulk helical DNLs and surface TFBs along the out-of-plane momentum direction.} (\textbf{a-n}), Energy contours measured at $\SI{-50}{\milli\electronvolt}$ at photon energies from $40$ to $\SI{105}{\electronvolt}$ to show the rotation of the Dirac node as indicated by pink arrows. The Dirac node is barely detectable at $80-\SI{95}{\electronvolt}$  due to suppression of the intensity by dipole matrix element effects. (\textbf{o}), Energy contours around the K point measured at $\SI{-100}{\milli\electronvolt}$ at photon energies from $\SI{40}{\electronvolt}$ (bottom) to $\SI{105}{\electronvolt}$ (top), which correspond to reduced $k_z$ values from $1.72c^*$ to $2.79c^*$. (\textbf{p}), Schematic drawing of helical DNL around the K$^\prime$ point, which exhibits the opposite chirality to that in the K point. (\textbf{q-v}), Dispersion images measured at photon energies from $\SI{40}{\electronvolt}$ to $\SI{105}{\electronvolt}$ to reveal the non-dispersive TFBs indicated by red arrows. All dispersion images were measured by cutting through the Dirac node and K point at each photon energy, as indicated in the insets.}
	\label{Fig.3}
\end{figure*}

Photon-energy dependent ARPES measurements have been performed to further confirm that the Dirac cone originates from the three-dimensional (3D) bulk bands and the TFBs correspond to the two-dimensional (2D) surface states (see more details in SI Appendix Fig.~S6). Figure~3a-n shows intensity contours measured with photon energies varying from $\SI{40}{\electronvolt}$ to $\SI{105}{\electronvolt}$, which is sufficient to cover an entire $k_z$ Brillouin zone. The bulk Dirac node is indicated by the pink dot in Fig.~3a, and the intensity tail from the bottom of the TFBs is indicated by the gray dot. While the bottom of the TFBs remains fixed at the K point, the Dirac node rotates around the K point in the $k_x$-$k_y$ plane when changing $k_z$, as indicated by the pink arrows in Fig.~3a-n. The rotation period for the Dirac node agrees with the periodicity of the $k_z$ Brillouin zone of RG (see SI Appendix Fig.~S7 for supporting information). Figure~3o shows the energy contours in the $k_x$-$k_y$-$k_z$ space, where the helix of the Dirac nodes in 3D is indicated by the pink curve. Moreover, ARPES intensity maps covering both K and K$^\prime$ valleys further show that the Dirac node helix around the K$^\prime$ point has the opposite chirality to that at the K point, as schematically illustrated by the blue curve in Fig.~3p (see SI Appendix Fig.~S8,9 for supporting experimental data and calculations at both K and K$^\prime$ valleys). Interestingly, in contrast to the $k_z$ dependent bulk Dirac helixes which rotate with $k_z$, the TFBs near $E_F$ are observed clearly at all photon energies (as indicated by red arrows in Fig.~3q-v) without noticeable $k_z$ dependence, confirming their 2D surface state origin.

The observation of surface TFBs and bulk helical DNLs demonstrates RG to be a topological helical Dirac nodal line semimetal, where Dirac crossing points form helixes in the three-dimensional momentum space \cite{FangZNLSCPB2016}. Here the TFBs form ``drumhead'' surface states, which are topologically protected by the bulk states. Moreover, the number of Dirac helixes is two, which is the minimum allowed by $\mathcal{PT}$ symmetry, and there are no additional bands near $E_F$ to complicate the electronic structure. Therefore, RG provides an ideal system for investigating the physics of helical Dirac nodal line semimetals with surface TFBs as well as electron-electron correlation.

\begin{figure*}
	\centering
	\includegraphics[width=17cm]{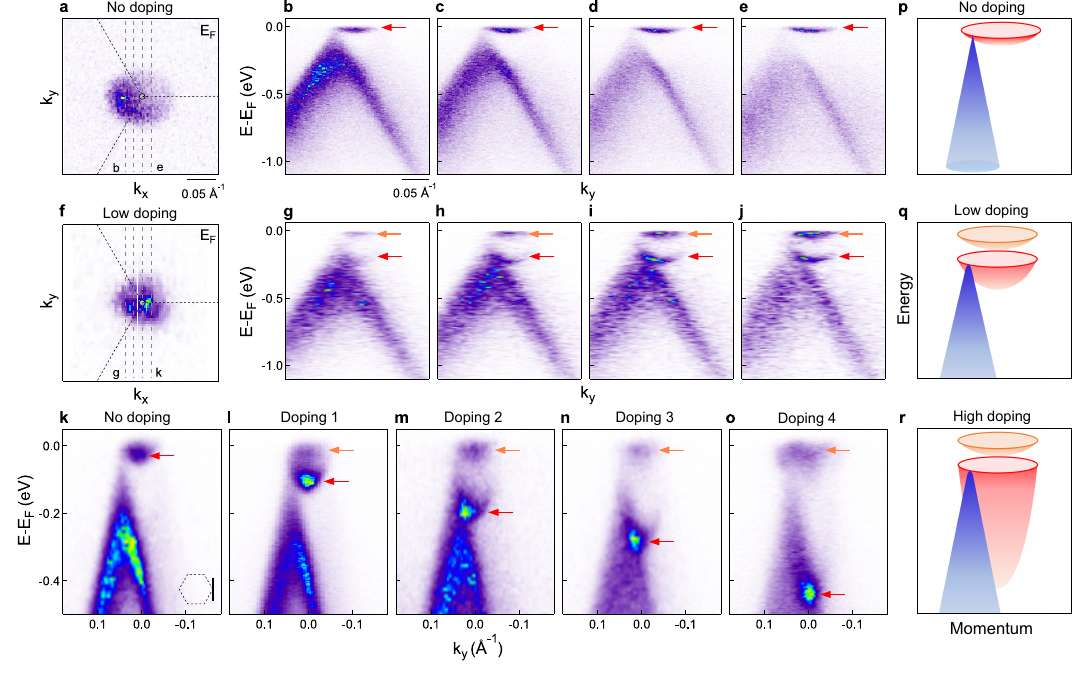}
	\caption{\textbf{Correlated evolution of the surface TFBs electronic structure upon \textbf{\emph{in situ}} electron doping.} (\textbf{a}), Fermi surface map measured before electron doping. (\textbf{b-e}), ARPES dispersion images cutting along the $k_y$ direction as indicated by gray broken lines in (\textbf{a}). (\textbf{f}), Fermi surface map measured at low electron doping. (\textbf{g-j}), ARPES dispersion images cutting along the $k_y$ direction as indicated by gray broken lines in (\textbf{f}). (\textbf{k-o}), ARPES Dispersion images cutting through the K point along $k_y$ direction upon electron doping. (\textbf{p-r}), Schematics for the undoped electronic structure, flavor splitting of TFBs, and an increase in bandwidth of the lower band upon electron doping, where orange and red colors indicate the upper and lower split TFBs respectively. The measurements here were performed on sample S1.}
	\label{fig4}
\end{figure*}

\section{Exotic evolution of the surface topological flat bands upon \textbf{\emph{in situ}} electron doping}

The compression of electrons into the surface TFBs within a small energy range suggests enhanced electron-electron correlation. Below we further investigate the effect of electron correlation upon {\it in situ} surface electron doping. Figure~4a-j shows a comparison of ARPES data measured at various momentum cuts on undoped (Fig.~4a-e) and slightly electron-doped (Fig.~4f-j) bulk RG. A splitting of the surface TFBs is clearly resolved upon electron doping (indicated by red and orange arrows in Fig.~4g-j). Since the ARPES probing depth of a few Angstrom is much smaller than the sample thickness of bulk RG, one cannot detect the surface TFBs from the bottom surface. We therefore conclude that the splitting is not caused by the lifted degeneracy between the top and bottom layers, but rather by other flavors, e.g. spin or valley spontaneous symmetry breaking, or correlation driven spectral weight splitting. Interestingly, while the upper band remains flat and pinned near $E_F$, the lower band becomes parabolic with an increase in the bandwidth (Fig.~4l-o). At the highest doping shown in Fig.~4o, the energy separation between the bottoms of the upper and lower bands reaches a maximum value of approximately $\SI{400}{\milli\electronvolt}$. Such a drastic increase in the bandwidth by nearly one order-of-magnitude as compared to the undoped case (schematically summarized in Fig.~4p-r) is an experimental signature of strong Coulomb repulsion when filling the TFBs.

\section{Hartree-Fock calculations suggest correlation effect beyond the mean-field theory}

In order to theoretically understand the doping evolution of the surface TFBs, we perform a self-consistent Hartree-Fock (HF) calculation (see SI Appendix for more details). We use a semi-infinite model of RG with a top layer and infinitely many layers below. The screened Coulomb interaction is projected into the four surface TFBs corresponding to the four spin-valley flavors, within the TFBs in-plane momentum radius $p_0$. The interaction between two electrons separated by a distance $r$ takes the Thomas-Fermi form $V(r)=\frac{q_e^2}{4\pi\epsilon_0 r}e^{-q_0r}$, where we set $q_0=\SI{0.1}{\per\angstrom}$, $q_e$ is the electron charge, and $\epsilon_0$ is the vacuum permittivity. The filling factor $\nu$ refers to the top surface TFBs confined within a momentum space (with a diameter of 0.1 $\AA^{-1}$) near the graphene Brillouin zone corner, which are the only states considered in the Hartree-Fock calculation (see more Hartree-Fock calculation details in the Supplementary Information). The band filling $\nu$ is defined to range from $\nu=-2$ (all surface TFBs empty) to $\nu=2$ (all surface TFBs occupied), with the TFB defined as the drumhead TFB within the momentum radius $p_0$. Figure~5 shows the evolution of the HF bands for fillings $-1.95\leq\nu\leq-1.15$, while results for larger fillings can be found in SI Appendix Fig.~S10. At generic partial fillings, we observe band splitting due to the spontaneous symmetry breaking of the 4-fold spin-valley degeneracy. Moreover, the bands become significantly dispersive due to the Coulomb interaction, with the energy difference between the band bottom and the Fermi level rising to the order of $\SI{400}{\milli\electronvolt}$, in agreement with the experimentally observed energy scale. We find that while the HF ground state is spin-valley polarized at all fillings, there are intervalley coherent states with very small excitation energy (approximately $10^{-3}$ or $10^{-2}~\si{\milli\electronvolt}$ per graphene unit cell). Therefore, the actual valley polarization may depend on factors not considered here, which is similar to what has been found in twisted bilayer graphene (TBG)\cite{Yazidani_Nature2023}.

\begin{figure*}[htbp]
	\centering
	\includegraphics[width=16.8cm]{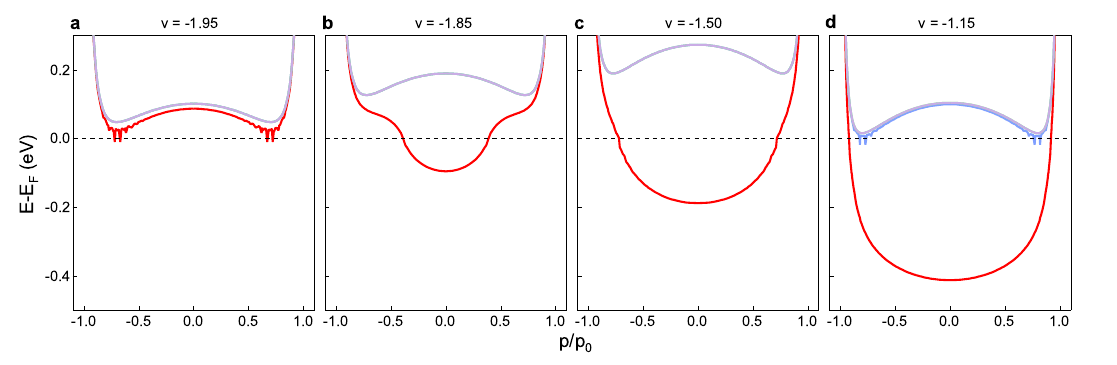}
	\caption{\textbf{Hartree-Fock band structures at negative fillings.} (\textbf{a-d}), Self-consistent Hartree-Fock calculations of the surface TFBs at fillings of $\nu = -1.95$, $-1.85$, $-1.50$, and $-1.15$, respectively. Each panel shows the four resulting Hartree-Fock bands. The top three bands are degenerate. The bottom of the lowest TFB (red curve) moves from near the Fermi level to approximately $-\SI{400}{\milli\electronvolt}$ (see SI Appendix Fig.~S10 for similar plots at fillings of $\nu = -1.0$, $-0.8$, $-0.4$, and $0.0$).}
	\label{Fig5}
\end{figure*}

Based on the comparison between HF calculation results (Fig.~5a-d) with experimental results (Fig.~4k-o) upon electron doping, we conjecture that the undoped RG in Fig.~4k is at a filling near $\nu = -2$ so that the only filled surface states are near the Fermi level with a narrow bandwidth. 
The experimentally observed slight filling of TFBs in the undoped graphite (Fig.~4k) could arise from a non-zero out-of-plane displacement field or possible doping from the substrate. As the surface electron doping increases, the lowest surface band shifts downwards and curves below the Fermi level (Fig.~5b-d), which may account for the lower band observed in Fig.~4k-o. However, our HF calculations cannot describe the upper band pinned near the Fermi level $E_F$, which is observed in Fig.~4k-o. We speculate that this pinned band originates from other factors such as correlation effects beyond the mean-field theory. For example, the surface flat bands and bulk dispersive bands here may resemble the topological Kondo model of twisted bilayer graphene (TBG) \cite{Song_PRL2022} consisting of flat bands and dispersive bands, which results in a Kondo resonance spectral weight peak at the Fermi level \cite{Daixi_PRB2022,Sarma_PRL2023,Bernevig_PRL2023,Song_PRB2024}, as observed experimentally in TBG \cite{YazdaniCascadeNat2020}. Further theoretical investigations are needed to establish a full understanding regarding the pinning of flat band upon doping. 

\section{Conclusion and outlook}
The discovery of topologically protected surface TFBs makes bulk RG a unique platform for bridging intriguing physics in various dimensionalities, such as the edge state of 1D zigzag graphene ribbons \cite{Steven_Nature2006,Tapaszto2014ZigzagFM}, 2D flat bands in moir\'e superlattices \cite{GuineaNRP2023}, and 3D nodal line semimetals with the 3D quantum Hall effect \cite{LinMFSSC2014}. The coexistence of nontrivial topological physics and correlation effects provides new opportunities for exploring intertwined topological physics and interaction-driven competing  symmetry-broken states, such as high-temperature surface superconductivity \cite{VolovikSCPRB2011}, topological superconductivity and chiral superconductivity \cite{SerbynPRB2023}.


		\begin{addendum}

		\item[Acknowledgement] We acknowledge useful discussions with Haizhou Lu, Jing Wang, and Yves H. Kwan. This work is supported by the National Key R$\&$D Program of China (Grant No.~2021YFA1400100), the National Natural Science Foundation of China (Grant No.~12234011, 52388201, 52025024, 92250305, 12327805, 11725418). H. Z. and C. B. acknowledge support from the Shuimu Tsinghua Scholar program, the Project funded by China Postdoctoral Science Foundation (Grant No.~2022M721887, 2022M721886), and the National Natural Science Foundation of China (Grant No.~12304226). M.S. and B.L. acknowledge support from National Science Foundation through Princeton University’s Materials Research Science and Engineering Center DMR-2011750, and the National Science Foundation under award DMR-2141966. K.W. and T.T. acknowledge support from the JSPS KAKENHI (Grant Numbers 20H00354 and 23H02052) and World Premier International Research Center Initiative (WPI), MEXT, Japan. This research used resources of the Advanced Light Source, which is a DOE Office of Science User Facility under contract No.~DE-AC02-05CH11231. Part of this research also used the Beamline 03U of the Shanghai Synchrotron Radiation Facility, which is supported by SiP$\cdot$ME$^2$ project under Contract No.~11227902 from National Natural Science Foundation of China. We acknowledge SOLEIL for the provision of synchrotron radiation facilities of beamline ANTARES.

		\item[Author Contributions] S.Z. conceived the research project. H.Z., Q.L., W.C., C.B., X.C., K.D., M.Y., J.A., P.D., J.D. and S.Z. performed the ARPES measurements and analyzed the ARPES data. H.Z., Q.L. prepared the samples. H.Z., Q.L. and P.Y. performed the AFM and Raman measurements. M.S., R.W., C.T., N.Z., J.L., M.L., P.T., Y.X., H.Y., B.L. and W.D. performed the calculations. K.W. and T.T. prepared BN crystals. H.Z. and S.Z. wrote the manuscript, and all authors commented on the manuscript.
		
		\item[Competing Interests] The authors declare that they have no competing interests.
		
		\item[Data availability] All relevant data of this study are available within the paper and its Supplementary Information files. Source data are provided with this paper.

	\end{addendum}

\end{document}